\documentclass[apj]{emulateapj}
\usepackage{graphics}
\usepackage{epsfig}
\usepackage{alltt}
\usepackage{verbatim}

\def\spose#1{\hbox to 0pt{#1\hss}}
\def\simlt{\mathrel{\spose{\lower 3pt\hbox{$\mathchar"218$}}
    \raise 2.0pt\hbox{$\mathchar"13C$}}}
\def\simgt{\mathrel{\spose{\lower 3pt\hbox{$\mathchar"218$}}
    \raise 2.0pt\hbox{$\mathchar"13E$}}}

\newcommand{\oiii}{\mbox{[\ion{O}{3}]} $\,$}
\newcommand{\oiiiw}{\mbox{[\ion{O}{3}] $\lambda$5007} $\,$}
\newcommand{\oiiiwn}{\mbox{[\ion{O}{3}] $\lambda$5007}}

\newcommand{\oiiiwwn}{\mbox{[\ion{O}{3}] $\lambda$4959}}

\newcommand{\hbn}{\mbox{H$\beta$}}
\newcommand{\ha}{\mbox{H$\alpha$} $\,$}
\newcommand{\han}{\mbox{H$\alpha$}}

\newcommand{\niin}{\mbox{[\ion{N}{2}] $\lambda$6584}}

\newcommand{\galaxy}{SDSS  J171544.05+600835.7 $\,$}
\newcommand{\galaxyn}{SDSS  J171544.05+600835.7}

\newcommand{\err}[2]{\ensuremath{^{_{+#1}}_{^{-#2}}}}
\newcommand{\ee}[2]{\ensuremath{{#1}\!\times\!10^{#2}}}
\newcommand{\ergcms}{\ensuremath{\mathrm{erg~cm}^{-2}~\mathrm{s}^{-1}}}
\newcommand{\chandra}{\textit{Chandra}}
\newcommand{\cxo}{\textit{Chandra X-ray Observatory}}

\newcommand{\pcmsq}{\mbox{cm$^{-2}$}}

\slugcomment{\it Submitted for Publication in ApJL}
\shortauthors{Comerford et al.}
\shorttitle{{\it Chandra} Observations of a 1.9 kpc Separation Double X-ray Source in a Candidate Dual AGN Galaxy}

\begin{document}

\title{Chandra Observations of a 1.9 kpc Separation Double X-ray Source \\ in a Candidate Dual AGN Galaxy at $\MakeLowercase{z}=0.16$}

\author{Julia M. Comerford\altaffilmark{1,3}, David Pooley\altaffilmark{1}, Brian F. Gerke\altaffilmark{2}, and Greg M. Madejski\altaffilmark{2}}

\thanks{$^3$W.J. McDonald Postdoctoral Fellow}

\affil{$^1$Astronomy Department, University of Texas at Austin, Austin, TX 78712}
\affil{$^2$Kavli Institute for Particle Astrophysics and Cosmology,
  M/S 29, Stanford Linear Accelerator Center, \\ 2575 Sand Hill Rd., 
  Menlo Park, CA 94725}
  
\begin{abstract}
We report {\it Chandra} observations of a double X-ray source in the $z=0.1569$ galaxy \galaxyn.  The galaxy was initially identified as a dual AGN candidate based on the double-peaked \oiiiw emission lines, with a line-of-sight velocity separation of 350 km s$^{-1}$, in its Sloan Digital Sky Survey spectrum.  We used the Kast Spectrograph at Lick Observatory to obtain two longslit spectra of the galaxy at two different position angles, which reveal that the two AGN emission components have not only a velocity offset, but also a projected spatial offset of 1.9 $h^{-1}_{70}$ kpc on the sky.  {\it Chandra}/ACIS observations of two X-ray sources with the same spatial offset and orientation as the optical emission suggest the galaxy most likely contains Compton-thick dual AGN, although the observations could also be explained by AGN jets. Deeper X-ray observations that reveal Fe K lines, if present, would distinguish between the two scenarios.  The observations of a double X-ray source in \galaxy are a proof of concept for a new, systematic detection method that selects promising dual AGN candidates from ground-based spectroscopy that exhibits both velocity and spatial offsets in the AGN emission features.
\end{abstract}


\section{Introduction}
\label{intro}

A wealth of observations have shown that galaxy mergers are common and that nearly all galaxies host a central supermassive black hole (SMBH). Consequently, some galaxies must host two SMBHs as the result of recent mergers.  These are known as dual SMBHs for the first $\sim 100$ Myr after the merger when they are at separations $\simgt 1$ kpc
\citep{BE80.1,MI01.1}. Dual-SMBH systems are an important testing ground for theories of galaxy formation and evolution.  For example, simulations predict that quasar feedback in mergers can have extreme effects on star formation \citep{SP05.1} and that the core-cusp division in nuclear stellar distributions may be caused by the scouring effects of dual SMBHs \citep{MI02.2,LA07.1}. A statistical study of dual SMBHs and their host galaxies would thus have important implications for theories of galaxy formation and  SMBH growth.

Dual SMBHs are observable when sufficient gas accretes onto them to power dual active galactic nuclei (AGN).  While there have been identifications of hundreds of binary quasar pairs at separations $>10$ kpc (e.g., \citealt{HE06.1,MY07.1,MY08.1,GR10.2}), as well as a handful of galaxy pairs where each galaxy hosts an AGN \citep{BA04.2,GU05.1,PI10.1}, very few AGN pairs have been observed in the next evolutionary stage where they coexist at kpc-scale separations in the same merger-remnant galaxy.  To date, the confirmed dual AGN were identified by radio or X-ray resolution of two AGN with separations of 7 kpc in the $z=0.02$ double radio source 3C 75 at the center of the galaxy cluster Abell 400 \citep{HU06.1}, 4 kpc in the $z=0.05$ ultraluminous infrared galaxy Mrk 463 \citep{BI08.1}, and 0.7 kpc in the $z=0.02$ ultraluminous infrared galaxy NGC 6240 \citep{KO03.1}.

In recent years dual AGN candidates have been selected as galaxies with double-peaked AGN emission lines, first in the DEEP2 Galaxy Redshift Survey \citep{GE07.2,CO09.1} and later in the Sloan Digital Sky Survey (SDSS; \citealt{WA09.1, LI10.1, SM10.1}).  Although a double-peaked line profile is expected for dual AGN, it could also be produced by gas kinematics in a single AGN; follow-up observations are necessary to distinguish between the two scenarios.  Follow-up optical longslit spectroscopy, near-infrared imaging, and adaptive optics imaging have added circumstantial evidence that many double-peaked systems may indeed be dual AGN \citep{FU10.1, LI10.2, SH10.1, GR11.1, RO11.1}, but direct resolution of two separate AGN is required for direct evidence of dual AGN.

\begin{figure}[!t]
\begin{center}
\vspace{-.5in}
\includegraphics[width=3.5in]{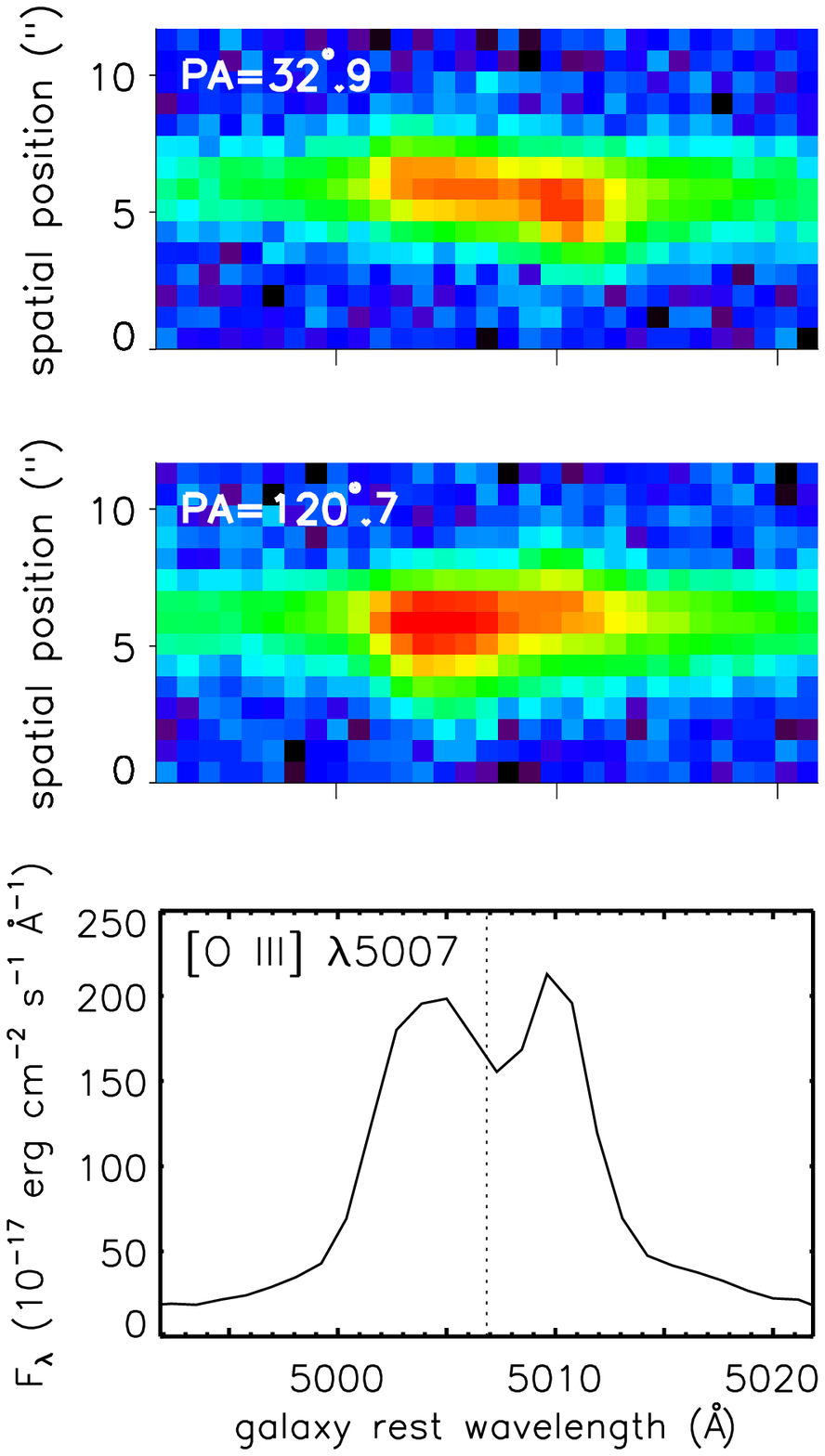}
\end{center}
\vspace{-.3in}
\caption{Segments of the two-dimensional Lick/Kast spectra at position angle $32^\circ.9$ east of north (top) and position angle $120^\circ.7$ east of north (middle) and the SDSS spectrum (bottom) for \galaxyn. Each spectrum is shifted to the rest frame of the host galaxy, and centered on the rest wavelength of \oiiiw (dotted vertical line).  In the Lick/Kast spectra, night-sky emission features have been subtracted and both vertical axes span $11\farcs7$ (31.8 $h^{-1}_{70}$ kpc at the $z=0.1569$ redshift of the galaxy). The double peaks in \oiiiw in the SDSS spectrum correspond to spatially-offset double emission features in the Lick/Kast spectra, suggesting the presence of dual AGN with a line-of-sight velocity separation of 350 km s$^{-1}$ and a projected spatial separation of 1.9 $h^{-1}_{70}$ kpc (or 0$\farcs$68) on the sky.  {\it Chandra} observations support the presence of dual AGN.}
\label{fig:spectra}
\end{figure}

Here we present such evidence for dual AGN in the form of {\it Chandra}/ACIS observations of two AGN sources separated by 1.9 $h^{-1}_{70}$ kpc in the $z=0.1569$ galaxy \galaxyn.   This galaxy was first identified as a dual AGN candidate by its double-peaked Type 2 AGN spectrum in SDSS, where the \oiiiw peaks are separated by 350 km s$^{-1}$ \citep{LI10.1, SM10.1}. Our follow-up Lick/Kast longslit spectra of the galaxy show two \oiiiw emission components separated on the sky by 1.9 $h^{-1}_{70}$ kpc, or 0$\farcs$68, an angular separation resolvable by {\it Chandra}.  The follow-up {\it Chandra} observations reveal double X-ray sources with the same spatial separation and orientation on the sky as the two \oiiiw emission components, indicating that \galaxy likely hosts Compton-thick dual AGN.  We assume a Hubble constant $H_0 =70 \,$ km s$^{-1}$ Mpc$^{-1}$,
$\Omega_m=0.3$, and $\Omega_\Lambda=0.7$ throughout, and all distances are given in physical (not comoving) units.

\section{Observations and Analysis}

\subsection{SDSS Spectrum}

The SDSS spectrum of \galaxy exhibits several double-peaked AGN emission lines, where one peak is blueshifted and one is redshifted relative to the systemic redshift of the host galaxy and the \oiiiw peaks have a line-of-sight velocity separation of 350 km s$^{-1}$ (\citealt{LI10.1, SM10.1}; Figure~\ref{fig:spectra}).  We fit two Gaussians each to the continuum-subtracted \oiiiwn, \hbn, \niin, and \ha line profiles and used the areas under the best-fit Gaussians as estimates of the line fluxes of the redshifted and blueshifted components of each line.  This yields line flux ratios of $\oiiiwn/\hbn=4.4 \pm 1.4$ and $\niin/\han=1.1 \pm 0.2$ for the blueshifted components, and $\oiiiwn/\hbn=8.6 \pm 3.1$ and $\niin/\han=0.5 \pm 0.1$ for the redshifted components, where the uncertainties are derived from propagating the errors in the parameters of the best-fit Gaussians.  These line flux ratios clearly indicate that both the redshifted and the blueshifted emission components are produced by AGN \citep{BA81.1, KE06.1}. 

\subsection{Lick/Kast Longslit Spectra}
\label{lick}

The SDSS fiber spectrum carries no spatial information about the source or sources of the emission producing the double-peaked lines.  Because this spatial information can help distinguish whether the emission is produced by two spatially-offset AGN or gas kinematics from a single AGN, we obtained follow-up slit spectroscopy of the galaxy.

We used the Kast Spectrograph on the Lick 3-m telescope to obtain spectra of the galaxy with a 1200 lines mm$^{-1}$ grating on UT 2009 August 17.  To determine the orientation of the AGN emission components on the plane of the sky, we observed the galaxy at two different position angles, $32^\circ.9$ east of north and $120^\circ.7$ east of north.  At each position angle we took three 1200 s exposures, and each spectrum spans the wavelength range 4790 -- 6200 \AA.  The data were reduced following standard procedures in IRAF and IDL.

The spectra at both position angles reveal two distinct emission components in \hbn, \oiiiwwn, and \oiiiw separated in both velocity and spatial position (Figure~\ref{fig:spectra}).
The \oiiiw emission has the highest signal-to-noise ratio, which enables the most precise separation measurements. For each spectrum we determine the projected spatial separation between the two \oiiiw emission features by measuring the spatial centroid of each emission component individually, then we combine the projected separation measurements at both position angles to determine the spatial separation and position angle on the sky (for details see Comerford et al., in prep.). We find the two \oiiiw emission components have a projected separation on the sky of $1.86 \pm 0.41$ $h^{-1}_{70}$ kpc, or 0$\farcs$684 $\pm$ 0$\farcs$151, and the position angle on the sky is $145^\circ.6$ east of north.

\begin{figure*}[!t]
\includegraphics[width=0.32\textwidth]{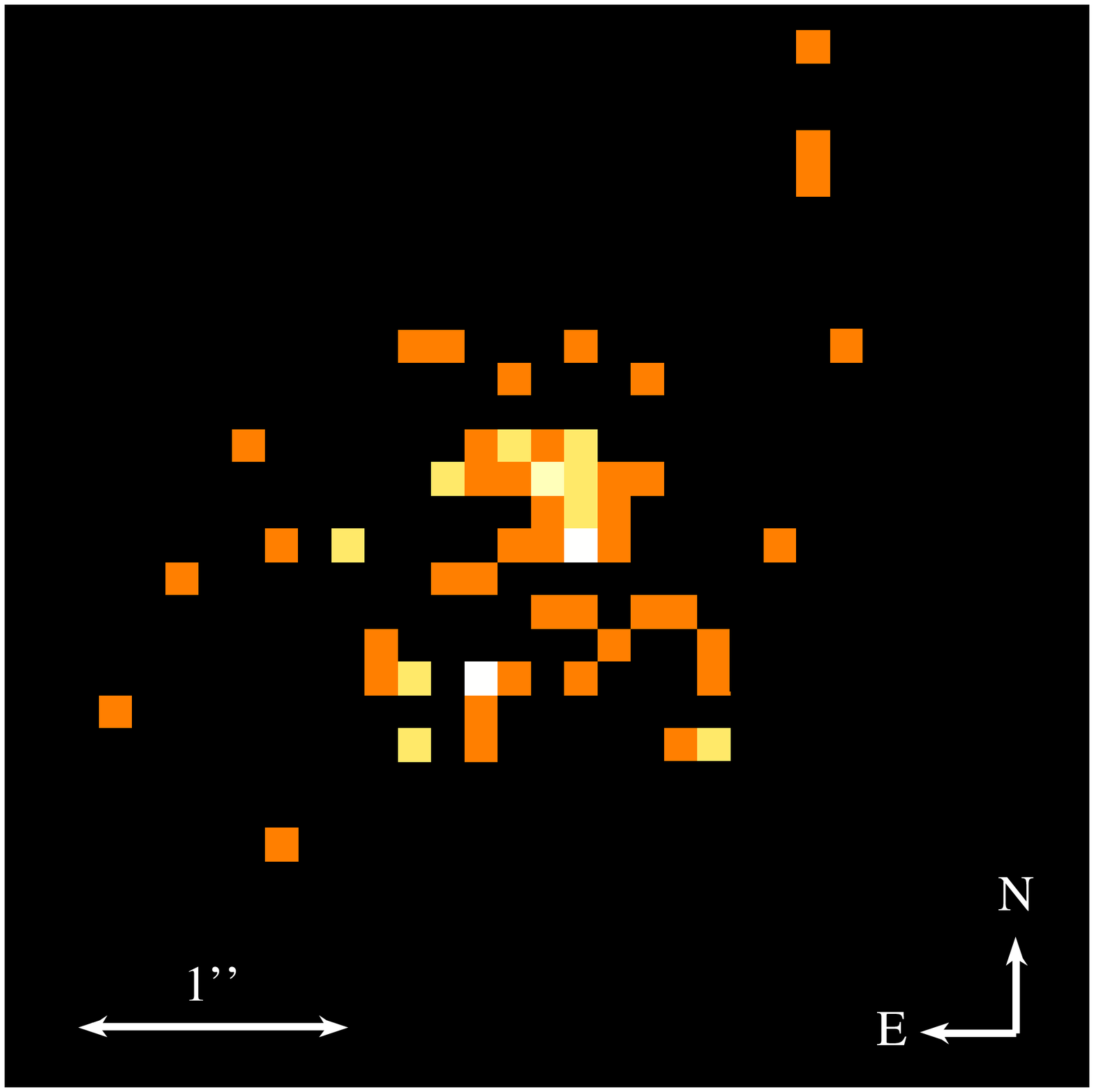}\hglue 0.02\textwidth
\includegraphics[width=0.32\textwidth]{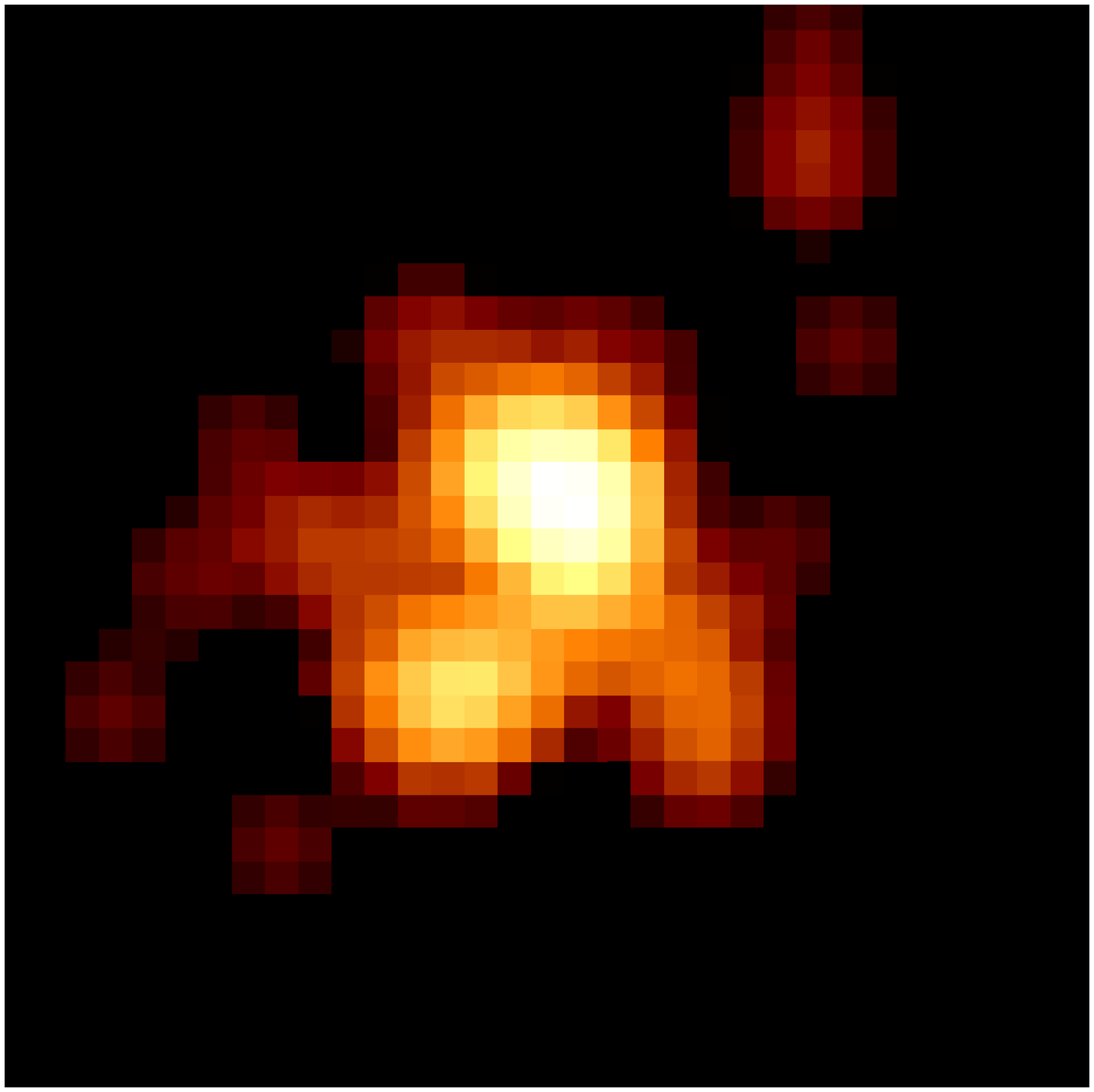}\hglue 0.02\textwidth
\includegraphics[width=0.32\textwidth]{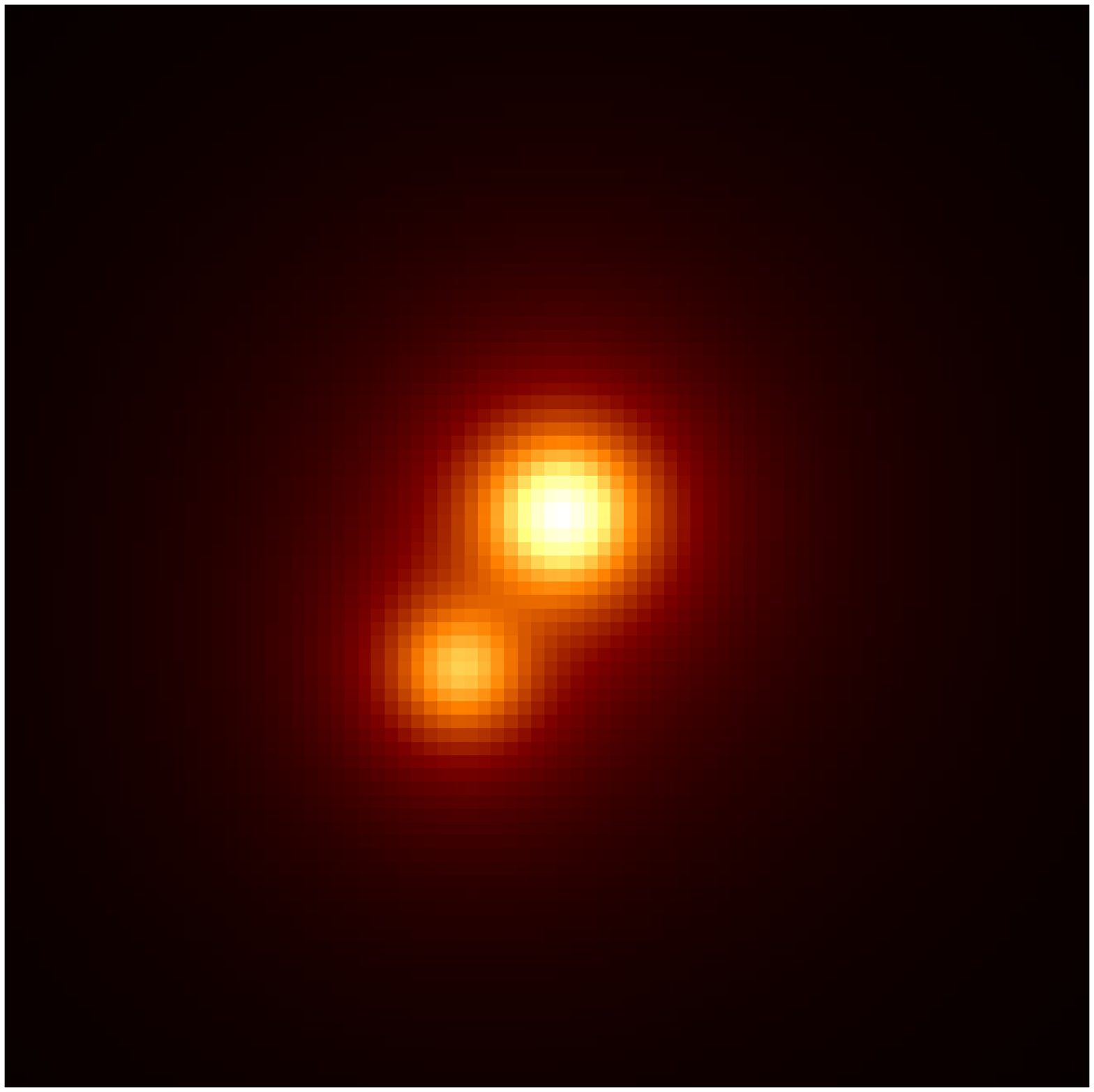}
\caption{{\it Chandra} X-ray image of the field of SDSS\,J171544.05+600835.7 using events in the 0.5--8 keV range (left), Gaussian-smoothed version of the data (center), and image of the two-component $\beta$-profile model (right).  All images are 4\arcsec\ on a side. The X-ray hardness ratios of the two sources are suggestive of Compton-thick dual AGN.}
\label{fig:xrayimage}
\end{figure*}

\subsection{Chandra Observations}
\label{chandra}

\galaxy was observed with the \cxo\ on 2011 March 17 beginning at 16:22 UT with an exposure time of 29\,669~s.  The observation was taken with the telescope aimpoint on the Advanced CCD Imaging Spectrometer (ACIS) S3 chip in ``timed exposure'' mode and telemetered to the ground in ``faint'' mode.  The observation was reduced using the latest \chandra\ software (CIAO\,4.3) and the most recent set of calibration files (CALDB\,4.4.1).  The data were reprocessed with the ``chandra\_repro'' script using the subpixel event repositioning algorithm of \cite{LI04.4}.  Intervals of strong background flaring were searched for, but none were found.

We made a sky image of the field of \galaxy at a resolution of 0\farcs0492 per pixel with events in the 0.3--8.0 keV energy range, and we fit two-dimensional models to that image.  All fits were performed in Sherpa \citep{FR01.2} using modified \citet{CA79.1} statistics (``cstat'' in Sherpa) and the \citet{NE65.1} optimization method (``simplex'' in Sherpa).  A family of fits was performed, using a grid of parameter starting points to ensure a proper sampling of the multidimensional fit space.

We fit a two-component source model with a fixed background (based on a source-free region near SDSS\,J171544.05+600835.7).  The source model was a $\beta$ profile, which is a two-dimensional Lorentzian with a varying power law of the form $I(r) = A(1+(r/r_0)^2)^{-\alpha}$ and is a good match to the \chandra\ PSF.  Based on other work \citep{PO09.1}, the power law index $\alpha$ was tied to the $r_0$ parameter, and both components were required to have the same $r_0$. Their positions and amplitudes were unconstrained.  The best fit amplitudes are 0.49\err{0.23}{0.16} and 0.21\err{0.12}{0.08} counts pixel$^{-1}$ for the northern and southern sources, respectively, and where the southern source is detected at 3.7$\sigma$.

The two X-ray components are separated by $1.85 \pm 0.22$ $h^{-1}_{70}$ kpc, or $0\farcs68\pm0\farcs08$, at a position angle of $147^\circ \pm 9^\circ$ east of north, where both the separation and the position angle are consistent with those measured for the two \oiiiw emission components in the Lick/Kast longslit spectra (\S~\ref{lick}).  The best fit model is shown in Figure~\ref{fig:xrayimage}.

To estimate the fluxes of the two components, we cannot reliably extract spectra of and make response files for the two separately.  We therefore extracted a spectrum of both sources together and use the results of our two-dimensional image fits to assign appropriate fractions of the total flux to each component.  We fit the unbinned spectrum in Sherpa using cstat statistics and the simplex method.  The spectral model was a simple absorbed power law with the column density constrained to be at least the Galactic value of $n_H=\ee{2.6}{20}~\pcmsq$ \citep{DI90.1}.  No additional absorption was preferred in the fit, and the best fit power law index was $1.9\pm0.2$.  The total unabsorbed 0.5--8~keV flux is \ee{1.5\pm0.4}{-14}~\ergcms.  The uncertainty was calculated using the ``sample\_energy\_flux'' tool in Sherpa, which takes into account uncertainties in all model parameters.  Using the results of our two-dimensional image fit, the northern component has a best-fit flux of $F_{0.5-8}=\ee{1.1}{-14}~\ergcms$, and the southern component has $F_{0.5-8}=\ee{4.4}{-15}~\ergcms$.

The 2--10~keV fluxes are $F_{2-10}=6.4 \pm 3.1 \times 10^{-15}~\ergcms$ for the northern component and $F_{2-10}=2.7 \pm 1.5 \times 10^{-15}~\ergcms$ for the southern component, which we compare to the 2--10~keV fluxes predicted from the \oiiiw fluxes using the scaling relation for Type 2 AGN in \cite{HE05.1}.  The predicted 2--10~keV fluxes are $3.3 \pm 7.9 \times 10^{-14}~\ergcms$ and $2.1 \pm 5.1 \times 10^{-14}~\ergcms$ for the redshifted and blueshifted components of \oiiiwn, respectively.  The measured 2--10~keV fluxes are hence a factor of several lower than but within the broad uncertainties of the predictions from \cite{HE05.1}.

Although we cannot fit separate spectra for the two components, we can extract the counts in small regions centered on them and form hardness ratios, defined as $\mathrm{HR}=(H-S)/(H+S)$ where $H$ is the number of counts in the 2--8 keV range and $S$ is the number of counts in the 0.5--2 keV range.  Counts were extracted from 0\farcs25 radius regions centered on each source, yielding 20 counts from the northern source with $\mathrm{HR}=-0.57\err{0.15}{0.19}$ and 10 counts from the southern source with $\mathrm{HR}=-0.37\err{0.24}{0.30}$.  Uncertainties on the hardness ratios were calculated using the Bayesian Estimation of Hardness Ratios package \citep{PA06.1}.  Judging from the measured hardness ratios, neither source is as hard as would be expected for moderately absorbed (but not Compton-thick) AGN, although the signal-to-noise ratio is very modest.

\begin{figure}[!t]
\hspace{-.5in}
\begin{center}
\includegraphics[width=2.5in]{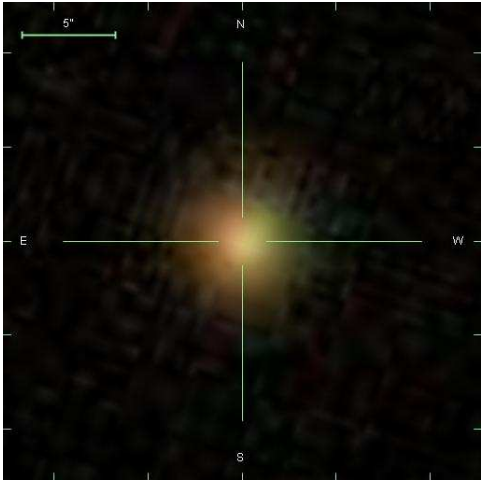}
\end{center}
\caption{SDSS image of \galaxyn.  Unlike the dual AGN host galaxies known to date, this galaxy has no indication of extreme star formation or an unusual morphology.  This image reveals no tidal features or companions, and no structure on scales $\simgt0\farcs1$ is reported from $K'$-band adaptive optics imaging \citep{FU10.1}.}
\label{fig:sdss}
\end{figure}

\subsection{Keck/LGSAO and SDSS Imaging}

If \galaxy has dual AGN separated by 1.9 $h^{-1}_{70}$ kpc that are the result of a galaxy merger, we might expect the two AGN to be coincident with two stellar components that are the remnants of the progenitor galaxies in the merger. In fact, \galaxy was one of 50 SDSS galaxies with double-peaked \oiii lines that was targeted for Keck II laser guide-star adaptive-optics (LGSAO) observations in \cite{FU10.1}, which found that 16 of these galaxies exhibited double stellar components with separations 0.6 -- 12 $h^{-1}_{70}$ kpc (or 0.2 -- 3$^{\prime\prime}$), suggestive of merging systems.  While \cite{FU10.1} do not show the image of \galaxyn, they report that the Keck/LGSAO $K'$-band image reveals only a single component with $K'=14.8$ and classify the galaxy as ``isolated".  The PSF FWHMs for their observations were $0\farcs065$ to $0\farcs130$, suggesting that \galaxy may have no substructure at $\simgt0\farcs1$ visible in these $K'$-band observations. 

Based on its SDSS photometry, \galaxy has a rest-frame $u-r$ color of 2.62 and an absolute $r$-band magnitude of -22.0, placing it on the red sequence of SDSS galaxies \citep{BA04.3}. The SDSS image of the galaxy shows no signatures of galaxy interaction (Figure~\ref{fig:sdss}).

\section{Interpretations}

Here we explore the physical mechanisms that could explain the observations of double AGN emission components and X-ray sources in \galaxyn.

The sources' high X-ray luminosities lead us to consider the possibility that they are ultra-luminous X-ray sources (ULXs), which are variable off-nuclear X-ray sources.  Some ULXs are also candidate intermediate mass black holes (e.g., \citealt{KA03.4, FA09.1}).   However, estimates of the black hole mass from the optical spectra place 
\galaxy in the supermassive black hole regime (Fu et al. 2010).  Further, the measured \oiii fluxes and luminosities are well above those typically measured in ULXs (e.g., \citealt{PO10.1, CS11.1}). 

Another possibility is that the galaxy hosts dual AGN that are the result of a triple SMBH interaction, where a gravitational slingshot effect \citep{SA74.1} ejected the least massive SMBH (corresponding to the southern AGN) while the remaining two SMBHs merged (producing the northern AGN).  The remaining stellar component associated with the southern AGN could be too faint to appear in the Keck/LGSAO observations. However, the optical observations suggest that both AGN have associated narrow-line regions and it is unclear how the ejected SMBH would maintain its narrow-line region.

We note also that if there is only a single source visible in the Keck/LGSAO image, a second stellar component could be either obscured or too faint to be within the detection threshold of the Keck/LGSAO observation.  If there are two AGN with associated stellar components, the stellar component accompanying the southern AGN is more likely to be obscured since it is harder and consequently more absorbed than the northern AGN. 
If the southern AGN is obscured and the northern AGN is not, the overall extinction for the system could be skewed towards the low value measured in \S~\ref{chandra}. 

Another explanation for our observations is that the galaxy hosts AGN jets that produce both \oiii and X-rays from a combination of photoionization from the AGN and collisional ionization from the jets (e.g., \citealt{ KR09.1, BI10.1}). For this scenario to explain our observations, either the AGN is completely obscured and only the jets are visible or one of the sources in our observations is the AGN while the other is the foreground jet and the background jet is obscured.  Both scenarios rely on significant obscuration in part of the galaxy, but this could be consistent with the low column density we measured if the rest of the galaxy is relatively unobscured. However, the jet scenario typically produces a bright core and a much fainter jet, whereas our observations show two sources that differ by only a factor of 2 in 2--10~keV luminosities.

The observations may also be explained by dual AGN that are Compton thick.  The X-ray spectra alone -- if confirmed -- exclude Compton-intermediate AGN, where the absorption would correspond to a column of $\sim 10^{22} - 10^{24}$ cm$^{-2}$, as there the observed X-ray flux would be dominated by partially (photoelectrically) absorbed continuum that is detectable as a hard X-ray spectrum. Instead the data suggest that both sources are either unabsorbed or the absorption is severe, essentially resulting in a Compton-thick spectrum.  The optical spectra seem to exclude classical, unabsorbed Type 1 AGN, so we are left with a Compton-thick scenario, with $\tau_{\rm Thomson}$ of at least a few (column $> 10^{25}$ cm$^{-2}$).  There, not only the primary nuclear soft X-ray flux is photoelectrically absorbed, but even the hard X-rays are suppressed by Compton opacity, and we detect only primary soft X-rays scattered back to our line of sight, as is the case in, e.g., NGC 1068 and other Compton-thick AGN. The measured 2--10~keV luminosities for the northern and southern components are within the range of values measured for Compton-thick AGN (e.g., \citealt{LE06.1}). The Compton-thick scenario is also consistent with the low column density of the galaxy, since the column density measurement presupposed that the source was not Compton thick.  

We conclude that SDSS J171544.05+600835.7 most likely hosts Compton-thick dual AGN, because it is the scenario that is most consistent with the existing data. AGN jets might also explain the observations, and deeper X-ray observations could distinguish between these two possibilities. 
These more sensitive X-ray measurements would enable a test of the Compton-thick scenario, since better spectral measurements in the soft X-ray band could provide a possible measurement of the Fe K line, which is seen in heavily absorbed AGN (for a discussion, see, e.g., \citealt{LE06.1}).  Sensitive hard X-ray measurements would provide much better constraints on the absorbing column.  {\it Hubble Space Telescope} narrow-band imaging of the \oiii emission could also show whether it has the biconical morphology expected for AGN jets.

\section{Conclusions}

We report observations of a double X-ray source with 1.9 $h^{-1}_{70}$ kpc, or 0$\farcs$68, projected spatial separation in the $z=0.1569$ candidate dual AGN galaxy \galaxyn.  This Seyfert 2 galaxy exhibits double-peaked \oiiiw emission lines with 350 km s$^{-1}$ line-of-sight velocity separation in its SDSS spectrum, and our follow-up Lick/Kast longslit spectra show two 1.9 $h^{-1}_{70}$ kpc separation \oiiiw emission components.  While the velocity and spatial offsets provide circumstantial evidence for dual AGN, these features could also be produced by gas kinematics from a single AGN.  The {\it Chandra} observations bolster the evidence for dual AGN, by revealing two X-ray components suggestive of Compton-thick AGN with the same spatial separation and orientation as the two sources of optical \oiiiw emission.

To date, dual AGN have typically been identified serendipitously because of the interesting characteristics of their host galaxies. These host galaxies include ultraluminous infrared galaxies \citep{KO03.1,BI08.1}, a double radio source at the center of a galaxy cluster \citep{HU06.1}, and a host galaxy with double bright nuclei and a tidal tail \citep{CO09.3}.  \galaxy is unlike these systems because there is nothing particularly noteworthy about the galaxy, which is a seemingly ordinary red sequence galaxy without tidal features visible in SDSS imaging or substructure reported in adaptive optics imaging.  Our observations suggest that dual AGN may be more ubiquitous and not limited to only galaxies with extreme star formation or unusual morphologies.

We have introduced a systematic, observational method for selecting promising dual AGN candidates, which have until now have only been identified through serendipitous discoveries of individual systems.  The method consists of three steps: 1) select dual AGN candidates as objects whose spectra exhibit double-peaked AGN emission lines in SDSS or other spectroscopic surveys of galaxies; 2) conduct follow-up longslit spectroscopy of the dual AGN candidates; 3) if the follow-up longslit spectra reveal an object has two spatially-distinct AGN emission components, use follow-up X-ray or radio observations to identify whether the object is a dual AGN. \galaxy is the first object for which this technique has been demonstrated, and our observations show it most likely hosts dual AGN; deeper X-ray observations would provide the definitive evidence.  Future observations will determine the general applicability of this systematic method for selecting dual AGN.

\acknowledgements J.M.C. acknowledges insightful discussions with Jenny Greene, as well as support from a W.J. McDonald Postdoctoral Fellowship.  The Texas Cosmology Center is supported by the College of Natural Sciences and the Department of Astronomy at the University of Texas at Austin and the McDonald Observatory.  B.F.G. and G.M.M. were supported by the U.S. Department of Energy under contract number DE-AC02-76SF00515.

\bibliographystyle{apj}

\begin{thebibliography}{47}
\expandafter\ifx\csname natexlab\endcsname\relax\def\natexlab#1{#1}\fi

\bibitem[{{Baldry} {et~al.}(2004){Baldry}, {Glazebrook}, {Brinkmann},
  {Ivezi{\'c}}, {Lupton}, {Nichol}, \& {Szalay}}]{BA04.3}
{Baldry}, I.~K., {Glazebrook}, K., {Brinkmann}, J., {Ivezi{\'c}}, {\v Z}.,
  {Lupton}, R.~H., {Nichol}, R.~C., \& {Szalay}, A.~S. 2004, \apj, 600, 681

\bibitem[{{Baldwin} {et~al.}(1981){Baldwin}, {Phillips}, \&
  {Terlevich}}]{BA81.1}
{Baldwin}, J.~A., {Phillips}, M.~M., \& {Terlevich}, R. 1981, \pasp, 93, 5

\bibitem[{{Ballo} {et~al.}(2004){Ballo}, {Braito}, {Della Ceca}, {Maraschi},
  {Tavecchio}, \& {Dadina}}]{BA04.2}
{Ballo}, L., {Braito}, V., {Della Ceca}, R., {Maraschi}, L., {Tavecchio}, F.,
  \& {Dadina}, M. 2004, \apj, 600, 634

\bibitem[{{Begelman} {et~al.}(1980){Begelman}, {Blandford}, \& {Rees}}]{BE80.1}
{Begelman}, M.~C., {Blandford}, R.~D., \& {Rees}, M.~J. 1980, \nat, 287, 307

\bibitem[{{Bianchi} {et~al.}(2010){Bianchi}, {Chiaberge}, {Evans}, {Guainazzi},
  {Baldi}, {Matt}, \& {Piconcelli}}]{BI10.1}
{Bianchi}, S., {Chiaberge}, M., {Evans}, D.~A., {Guainazzi}, M., {Baldi},
  R.~D., {Matt}, G., \& {Piconcelli}, E. 2010, \mnras, 405, 553

\bibitem[{{Bianchi} {et~al.}(2008){Bianchi}, {Chiaberge}, {Piconcelli},
  {Guainazzi}, \& {Matt}}]{BI08.1}
{Bianchi}, S., {Chiaberge}, M., {Piconcelli}, E., {Guainazzi}, M., \& {Matt},
  G. 2008, \mnras, 386, 105

\bibitem[{{Cash}(1979)}]{CA79.1}
{Cash}, W. 1979, \apj, 228, 939

\bibitem[{{Comerford} {et~al.}(2009{\natexlab{a}}){Comerford}, {Gerke},
  {Newman}, {Davis}, {Yan}, {Cooper}, {Faber}, {Koo}, {Coil}, {Rosario}, \&
  {Dutton}}]{CO09.1}
{Comerford}, J.~M., {Gerke}, B.~F., {Newman}, J.~A., {Davis}, M., {Yan}, R.,
  {Cooper}, M.~C., {Faber}, S.~M., {Koo}, D.~C., {Coil}, A.~L., {Rosario},
  D.~J., \& {Dutton}, A.~A. 2009{\natexlab{a}}, \apj, 698, 956

\bibitem[{{Comerford} {et~al.}(2009{\natexlab{b}}){Comerford}, {Griffith},
  {Gerke}, {Cooper}, {Newman}, {Davis}, \& {Stern}}]{CO09.3}
{Comerford}, J.~M., {Griffith}, R.~L., {Gerke}, B.~F., {Cooper}, M.~C.,
  {Newman}, J.~A., {Davis}, M., \& {Stern}, D. 2009{\natexlab{b}}, \apjl, 702,
  L82

\bibitem[{{Cseh} {et~al.}(2011){Cseh}, {Gris{\'e}}, {Corbel}, \&
  {Kaaret}}]{CS11.1}
{Cseh}, D., {Gris{\'e}}, F., {Corbel}, S., \& {Kaaret}, P. 2011, \apjl, 728,
  L5+

\bibitem[{{Dickey} \& {Lockman}(1990)}]{DI90.1}
{Dickey}, J.~M., \& {Lockman}, F.~J. 1990, \araa, 28, 215

\bibitem[{{Farrell} {et~al.}(2009){Farrell}, {Webb}, {Barret}, {Godet}, \&
  {Rodrigues}}]{FA09.1}
{Farrell}, S.~A., {Webb}, N.~A., {Barret}, D., {Godet}, O., \& {Rodrigues},
  J.~M. 2009, \nat, 460, 73

\bibitem[{{Freeman} {et~al.}(2001){Freeman}, {Doe}, \&
  {Siemiginowska}}]{FR01.2}
{Freeman}, P., {Doe}, S., \& {Siemiginowska}, A. 2001, in Society of
  Photo-Optical Instrumentation Engineers (SPIE) Conference Series, Vol. 4477,
  Society of Photo-Optical Instrumentation Engineers (SPIE) Conference Series,
  ed. {J.-L.~Starck \& F.~D.~Murtagh}, 76--87

\bibitem[{{Fu} {et~al.}(2010){Fu}, {Myers}, {Djorgovski}, \& {Yan}}]{FU10.1}
{Fu}, H., {Myers}, A.~D., {Djorgovski}, S.~G., \& {Yan}, L. 2010, ArXiv
  e-prints

\bibitem[{{Gerke} {et~al.}(2007)}]{GE07.2}
{Gerke}, B.~F., {et~al.} 2007, \apjl, 660, L23

\bibitem[{{Green} {et~al.}(2010){Green}, {Myers}, {Barkhouse}, {Mulchaey},
  {Bennert}, {Cox}, \& {Aldcroft}}]{GR10.2}
{Green}, P.~J., {Myers}, A.~D., {Barkhouse}, W.~A., {Mulchaey}, J.~S.,
  {Bennert}, V.~N., {Cox}, T.~J., \& {Aldcroft}, T.~L. 2010, \apj, 710, 1578

\bibitem[{{Greene} {et~al.}(2011){Greene}, {Zakamska}, {Ho}, \&
  {Barth}}]{GR11.1}
{Greene}, J.~E., {Zakamska}, N.~L., {Ho}, L.~C., \& {Barth}, A.~J. 2011, \apj,
  732, 9

\bibitem[{{Guainazzi} {et~al.}(2005){Guainazzi}, {Piconcelli},
  {Jim{\'e}nez-Bail{\'o}n}, \& {Matt}}]{GU05.1}
{Guainazzi}, M., {Piconcelli}, E., {Jim{\'e}nez-Bail{\'o}n}, E., \& {Matt}, G.
  2005, \aap, 429, L9

\bibitem[{{Heckman} {et~al.}(2005){Heckman}, {Ptak}, {Hornschemeier}, \&
  {Kauffmann}}]{HE05.1}
{Heckman}, T.~M., {Ptak}, A., {Hornschemeier}, A., \& {Kauffmann}, G. 2005,
  \apj, 634, 161

\bibitem[{{Hennawi} {et~al.}(2006){Hennawi}, {Strauss}, {Oguri}, {Inada},
  {Richards}, {Pindor}, {Schneider}, {Becker}, {Gregg}, {Hall}, {Johnston},
  {Fan}, {Burles}, {Schlegel}, {Gunn}, {Lupton}, {Bahcall}, {Brunner}, \&
  {Brinkmann}}]{HE06.1}
{Hennawi}, J.~F., {Strauss}, M.~A., {Oguri}, M., {Inada}, N., {Richards},
  G.~T., {Pindor}, B., {Schneider}, D.~P., {Becker}, R.~H., {Gregg}, M.~D.,
  {Hall}, P.~B., {Johnston}, D.~E., {Fan}, X., {Burles}, S., {Schlegel}, D.~J.,
  {Gunn}, J.~E., {Lupton}, R.~H., {Bahcall}, N.~A., {Brunner}, R.~J., \&
  {Brinkmann}, J. 2006, \aj, 131, 1

\bibitem[{{Hudson} {et~al.}(2006){Hudson}, {Reiprich}, {Clarke}, \&
  {Sarazin}}]{HU06.1}
{Hudson}, D.~S., {Reiprich}, T.~H., {Clarke}, T.~E., \& {Sarazin}, C.~L. 2006,
  \aap, 453, 433

\bibitem[{{Kaaret} {et~al.}(2003){Kaaret}, {Corbel}, {Prestwich}, \&
  {Zezas}}]{KA03.4}
{Kaaret}, P., {Corbel}, S., {Prestwich}, A.~H., \& {Zezas}, A. 2003, Science,
  299, 365

\bibitem[{{Kewley} {et~al.}(2006){Kewley}, {Groves}, {Kauffmann}, \&
  {Heckman}}]{KE06.1}
{Kewley}, L.~J., {Groves}, B., {Kauffmann}, G., \& {Heckman}, T. 2006, \mnras,
  372, 961

\bibitem[{{Komossa} {et~al.}(2003){Komossa}, {Burwitz}, {Hasinger}, {Predehl},
  {Kaastra}, \& {Ikebe}}]{KO03.1}
{Komossa}, S., {Burwitz}, V., {Hasinger}, G., {Predehl}, P., {Kaastra}, J.~S.,
  \& {Ikebe}, Y. 2003, \apjl, 582, L15

\bibitem[{{Kraemer} {et~al.}(2009){Kraemer}, {Trippe}, {Crenshaw},
  {Mel{\'e}ndez}, {Schmitt}, \& {Fischer}}]{KR09.1}
{Kraemer}, S.~B., {Trippe}, M.~L., {Crenshaw}, D.~M., {Mel{\'e}ndez}, M.,
  {Schmitt}, H.~R., \& {Fischer}, T.~C. 2009, \apj, 698, 106

\bibitem[{{Lauer} {et~al.}(2007){Lauer}, {Faber}, {Richstone}, {Gebhardt},
  {Tremaine}, {Postman}, {Dressler}, {Aller}, {Filippenko}, {Green}, {Ho},
  {Kormendy}, {Magorrian}, \& {Pinkney}}]{LA07.1}
{Lauer}, T.~R., {Faber}, S.~M., {Richstone}, D., {Gebhardt}, K., {Tremaine},
  S., {Postman}, M., {Dressler}, A., {Aller}, M.~C., {Filippenko}, A.~V.,
  {Green}, R., {Ho}, L.~C., {Kormendy}, J., {Magorrian}, J., \& {Pinkney}, J.
  2007, \apj, 662, 808

\bibitem[{{Levenson} {et~al.}(2006){Levenson}, {Heckman}, {Krolik}, {Weaver},
  \& {{\.Z}ycki}}]{LE06.1}
{Levenson}, N.~A., {Heckman}, T.~M., {Krolik}, J.~H., {Weaver}, K.~A., \&
  {{\.Z}ycki}, P.~T. 2006, \apj, 648, 111

\bibitem[{{Li} {et~al.}(2004){Li}, {Kastner}, {Prigozhin}, {Schulz},
  {Feigelson}, \& {Getman}}]{LI04.4}
{Li}, J., {Kastner}, J.~H., {Prigozhin}, G.~Y., {Schulz}, N.~S., {Feigelson},
  E.~D., \& {Getman}, K.~V. 2004, \apj, 610, 1204

\bibitem[{{Liu} {et~al.}(2010{\natexlab{a}}){Liu}, {Greene}, {Shen}, \&
  {Strauss}}]{LI10.2}
{Liu}, X., {Greene}, J.~E., {Shen}, Y., \& {Strauss}, M.~A. 2010{\natexlab{a}},
  \apjl, 715, L30

\bibitem[{{Liu} {et~al.}(2010{\natexlab{b}}){Liu}, {Shen}, {Strauss}, \&
  {Greene}}]{LI10.1}
{Liu}, X., {Shen}, Y., {Strauss}, M.~A., \& {Greene}, J.~E. 2010{\natexlab{b}},
  \apj, 708, 427

\bibitem[{{Milosavljevi{\'c}} \& {Merritt}(2001)}]{MI01.1}
{Milosavljevi{\'c}}, M., \& {Merritt}, D. 2001, \apj, 563, 34

\bibitem[{{Milosavljevi{\'c}} {et~al.}(2002){Milosavljevi{\'c}}, {Merritt},
  {Rest}, \& {van den Bosch}}]{MI02.2}
{Milosavljevi{\'c}}, M., {Merritt}, D., {Rest}, A., \& {van den Bosch}, F.~C.
  2002, \mnras, 331, L51

\bibitem[{{Myers} {et~al.}(2007){Myers}, {Brunner}, {Richards}, {Nichol},
  {Schneider}, \& {Bahcall}}]{MY07.1}
{Myers}, A.~D., {Brunner}, R.~J., {Richards}, G.~T., {Nichol}, R.~C.,
  {Schneider}, D.~P., \& {Bahcall}, N.~A. 2007, \apj, 658, 99

\bibitem[{{Myers} {et~al.}(2008){Myers}, {Richards}, {Brunner}, {Schneider},
  {Strand}, {Hall}, {Blomquist}, \& {York}}]{MY08.1}
{Myers}, A.~D., {Richards}, G.~T., {Brunner}, R.~J., {Schneider}, D.~P.,
  {Strand}, N.~E., {Hall}, P.~B., {Blomquist}, J.~A., \& {York}, D.~G. 2008,
  \apj, 678, 635

\bibitem[{{Nelder} \& {Mead}(1965)}]{NE65.1}
{Nelder}, J.~A., \& {Mead}, R. 1965, The Computer Journal, 7, 308

\bibitem[{{Park} {et~al.}(2006){Park}, {Kashyap}, {Siemiginowska}, {van Dyk},
  {Zezas}, {Heinke}, \& {Wargelin}}]{PA06.1}
{Park}, T., {Kashyap}, V.~L., {Siemiginowska}, A., {van Dyk}, D.~A., {Zezas},
  A., {Heinke}, C., \& {Wargelin}, B.~J. 2006, \apj, 652, 610

\bibitem[{{Piconcelli} {et~al.}(2010){Piconcelli}, {Vignali}, {Bianchi},
  {Mathur}, {Fiore}, {Guainazzi}, {Lanzuisi}, {Maiolino}, \&
  {Nicastro}}]{PI10.1}
{Piconcelli}, E., {Vignali}, C., {Bianchi}, S., {Mathur}, S., {Fiore}, F.,
  {Guainazzi}, M., {Lanzuisi}, G., {Maiolino}, R., \& {Nicastro}, F. 2010,
  \apjl, 722, L147

\bibitem[{{Pooley} {et~al.}(2009){Pooley}, {Rappaport}, {Blackburne},
  {Schechter}, {Schwab}, \& {Wambsganss}}]{PO09.1}
{Pooley}, D., {Rappaport}, S., {Blackburne}, J., {Schechter}, P.~L., {Schwab},
  J., \& {Wambsganss}, J. 2009, \apj, 697, 1892

\bibitem[{{Porter}(2010)}]{PO10.1}
{Porter}, R.~L. 2010, \mnras, 407, L59

\bibitem[{{Rosario} {et~al.}(2011){Rosario}, {McGurk}, {Max}, {Shields}, \&
  {Smith}}]{RO11.1}
{Rosario}, D.~J., {McGurk}, R.~C., {Max}, C.~E., {Shields}, G.~A., \& {Smith},
  K.~L. 2011, ArXiv e-prints

\bibitem[{{Saslaw} {et~al.}(1974){Saslaw}, {Valtonen}, \& {Aarseth}}]{SA74.1}
{Saslaw}, W.~C., {Valtonen}, M.~J., \& {Aarseth}, S.~J. 1974, \apj, 190, 253

\bibitem[{{Shen} {et~al.}(2010){Shen}, {Liu}, {Greene}, \& {Strauss}}]{SH10.1}
{Shen}, Y., {Liu}, X., {Greene}, J., \& {Strauss}, M. 2010, ArXiv e-prints

\bibitem[{{Smith} {et~al.}(2010){Smith}, {Shields}, {Bonning}, {McMullen},
  {Rosario}, \& {Salviander}}]{SM10.1}
{Smith}, K.~L., {Shields}, G.~A., {Bonning}, E.~W., {McMullen}, C.~C.,
  {Rosario}, D.~J., \& {Salviander}, S. 2010, \apj, 716, 866

\bibitem[{{Springel} {et~al.}(2005){Springel}, {Di Matteo}, \&
  {Hernquist}}]{SP05.1}
{Springel}, V., {Di Matteo}, T., \& {Hernquist}, L. 2005, \mnras, 361, 776

\bibitem[{{Wang} {et~al.}(2009){Wang}, {Chen}, {Hu}, {Mao}, {Zhang}, \&
  {Bian}}]{WA09.1}
{Wang}, J., {Chen}, Y., {Hu}, C., {Mao}, W., {Zhang}, S., \& {Bian}, W. 2009,
  \apjl, 705, L76

\end{thebibliography}

\end{document}